\documentclass[onecolumn,aps,floats,superscriptaddress,showpacs]{revtex4}
\usepackage{amsmath}

\usepackage{epsfig}
\usepackage{graphicx}
\usepackage{dcolumn}
\usepackage{bm}

\newcommand{\beq}{\begin{equation}}
\newcommand{\eeq}{\end{equation}}
\newcommand{\beqa}{\begin{eqnarray}}
\newcommand{\eeqa}{\end{eqnarray}}

\def\gapp{\lower.35em\hbox{$\stackrel{\textstyle>}{\sim}$}}
\def\lapp{\lower.35em\hbox{$\stackrel{\textstyle<}{\sim}$}}

\begin{document}
\bibliographystyle{apsrev}

\title{Renormalization of Coulomb interaction in graphene: Computing observable quantities}

\author{ Fernando de Juan and Adolfo G. Grushin and Mar\'{\i}a A. H. Vozmediano}

\affiliation{Instituto de Ciencia de Materiales de Madrid,\\
CSIC, Cantoblanco, E-28049 Madrid, Spain.}

\date{\today}
\begin{abstract}
We  address the computation of physical observables in graphene
in the presence of Coulomb interactions of density-density type
modeled with a static  Coulomb potential within a quantum field theory
perturbative renormalization scheme. We show that all the
divergences encountered in the physical quantities are associated
to the one loop electron self-energy and can be determined without
ambiguities by a proper renormalization of the Fermi velocity. The
renormalization of the photon polarization to second order in perturbation theory
- a quantity  directly related to the optical conductivity - is
given as an example.

\end{abstract}
%
%
%
%
\maketitle
\section{Introduction}
\label{intro} The role of many body corrections to the physics
of graphene is at this point uncertain. While in the first
transport experiments electron-electron interactions seemed not
to play a major role \cite{Netal04,Netal05,Zetal05} more recent
measurements \cite{Jetal07,Oetal07,ZSetal08,LLA09,NBetal08,Letal08} and the observation of the fractional Quantum Hall effect \cite{DSetal09,BGetal09} indicate the possible importance of the Coulomb interaction to the intrinsic properties of graphene.

The problem of the nature of the interacting system was addressed
in detail in the early works on graphene both in the weak
\cite{GGV94,GGV96,GGV99,GGV01} and in the strong coupling
limit\cite{K01,K01b,GMS02}. The main issue was to determine the
infrared nature of the system --its Fermi liquid character and its
associated physical properties--. In the weak coupling regime
\cite{GGV94} with a perturbative renormalization group (RG) analysis it was shown that intrinsic graphene behaves as a
strange Fermi liquid in the sense that all possible interactions
are marginally irrelevant but the inverse electron lifetime grows
linearly with the energy instead of quadratically \cite{GGV96}. An
upward renormalization of the Fermi velocity at low energies was
also predicted what in turn implied a downward renormalization of
the effective Coulomb interaction ($g=\frac{e^2}{ 4\pi \epsilon v}$) to
the infrared. Experimental indications of both the linear inverse
lifetime of the electron \cite{LLA09,Jetal07} and of the Fermi
velocity dependence with the energy \cite{Letal08} have been
recently reported. 
The results of these early works have been reproduced and pushed
forward under several approaches after the synthesis of graphene
\cite{DS06a,Macetal07,Mi07,Mi08,KUC08}. 

A very important aspect of renormalization was left aside in these previous works: the renormalization of the parameters of the theory that allow to determine the physical observables. We will here follow the standard quantum field theory (QFT) approach to renormalization
aiming to 
give a prescription
to calculate unambiguous, cutoff-independent observable results to any order
in perturbation theory. We show that a proper renormalization of the Fermi velocity renders all physical quantities finite and unambiguous to second order in
perturbation theory. We work out as an example the renormalization of the photon self-energy, a quantity directly related to the Coulomb interaction corrections to the optical properties of the system \cite{Mi07,HJV08,Mi08,SS09,GVV09}. 

\section{ Renormalization in Quantum Field Theory}
\label{ren}

Ultraviolet divergences arise in QFT due to the
singular behavior of the fields at very short distances in real
space - or at very large energies in Fourier space-. 
Renormalization \cite{N78,C84} is a prescription to
get rid of ultraviolet divergences and construct sensible models
where physical quantities can be accurately computed. The idea is
that the ultraviolet divergences can be canceled by adding counter terms to the
Lagrangian what amounts to a
redefinition of the parameters (mass, coupling constant, wave
function) of the theory. The process is usually done order by order in
perturbation theory. If done appropriately at the end one
finds finite results independent of the regularization procedure 
in the computation of physical observables. The prize to pay in this process is the necessity to fix the values of the renormalized parameters at a given energy
from some well-chosen experimental inputs. This "renormalization
prescription" is crucial in the process and the basis of the later
renormalization group (RG) developments in QFT.

The deduction of the experimental values of the parameters from a given experimental
measure often involves phenomenological
assumptions. Very good examples of these difficulties are provided
by the interplay between the accurate determination of the fine
structure constant of QED - that often is done from solid state
measurements of the electron precession in magnetic fields - and
its feedback  to determine the anomalous magnetic moment of the
electron with the actual precision of better than one part in a 
trillion \cite{HFG08}.

In the condensed matter applications the difficulty increases due
to the fact that we do not have ``scattering" experiments involving
the asymptotic states of the fields but transport measurements
that are influenced by all kinds of extrinsic factors (disorder,
doping, substrate). Yet the renormalization program can be adapted to condensed  matter systems that admit a continuum effective description, graphene been one of the best examples. The independence of the observable quantities on the cutoff guaranteed by the procedure makes it irrelevant whether or not the cutoff has to be taken to infinity or to a finite value defined at high energies as the inverse lattice spacing. It ensures that the observables of the effective low-energy theory do depend on the high energy  only through the renormalization of the effective parameters and do not have an explicit dependence on high energy quantities.

Next we will see these words at work in the concrete case of
graphene physics. The program consists of identifying the
primitively divergent Feynman graphs, adding counterterms to
subtract the divergences, redefine the parameters of the model to
absorb the infinities, and fix the finite parts of the vertex
correlation functions by a renormalization prescription. For this
last step we need a number of external conditions (observable
data) --the renormalization conditions-- equal to the number of
parameters to be renormalized. The original RG equations in the
QFT approach were established to demonstrate the independence of
the observable quantities on the renormalization prescription. Any
two set of experimental data will give rise to the same result for
a cross section.

\section{ Renormalization of the graphene model}
\label{renG}
\subsection{Definition of the model}
\begin{figure}
\begin{center}
\includegraphics[scale=0.3]{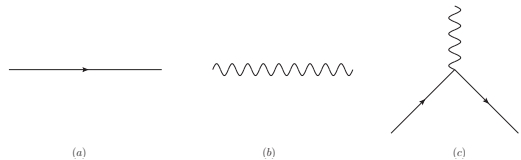}
\caption{Tree level Feynman diagrams (see text for details): (a)
Electron propagator (b) Photon propagator (c) Interaction vertex.}
\label{fig:Feynrules}
\end{center}
\end{figure}

It is known that the low energy excitations of a graphene sheet
around a Fermi point are well described by the massless Dirac
equation in two spacial dimensions \cite{W47,SW58}. The density of
states at the Fermi level vanishes and the Coulomb interactions
are unscreened \cite{GGV94}. The non-interacting model considering
a single Fermi point is described by the Hamiltonian
\begin{equation}
{\cal H}= \hbar v_F \int d^2 {\bf r} \bar{\psi}({\bf r})
\gamma^i\partial_i \psi ({\bf r})\;, \label{freeH}
\end{equation}
where $i=1,2$, $\bar\psi({\bf r})=\psi^+({\bf r})\gamma^0$, and
the gamma matrices can be chosen as $\gamma_x=\sigma_2,
\gamma_y=-\sigma_1, \gamma^0=\sigma_3 $ . $\sigma_i$ are the Pauli matrices and
$v_{F}$ is the Fermi velocity (to be defined unambiguosly in the next section). 

We will follow ref. \cite{GGV99} and model the electron-electron
interaction as a density-density interaction mediated by a scalar
potential. The (instantaneous) Coulomb interaction can be
described by the scalar component of the gauge field:
\begin{equation}
{\cal H}_{\rm int}= e\int d^2 {\bf r} \bar{\psi}({\bf r})\gamma^0
\psi ({\bf r})A_0({\bf r})\;, \label{hscalar}
\end{equation}
with our choice of gamma matrices. In order to define the renormalized theory we need a Lagrangian, 
a renormalization scheme (i.e. a regularization method and a set of renormalization conditions), 
and the experimentally measured parameters associated with these conditions. 

The Lagrangian is
\begin{equation}
\mathcal{L}=\int d^3k \bar{\psi}\left[\gamma^{0}k_0 + v\bm{\gamma}\cdot\mathbf{k}\right]\psi - e\bar{\psi}\gamma^{0}\psi A_{0}.
\label{lg1}
\end{equation}
The Lagrangian (\ref{lg1}) contains four quantities that can be redefined: the velocity $v$, the parameter $e$ in the interaction, and the electron and gauge field wave functions. 

The electron and photon
propagators in momentum space Fig. \ref{fig:Feynrules} (a) and (b) are given by
\begin{equation}
G_0(k^0, {\bf k})=i\frac{\gamma^0 k_0+ v\bm{\gamma} \cdot {\bf
k}}{-(k^0)^2+v^2{\bf k}^2}\;,
\end{equation}
\begin{equation}
\label{barephoton}
\Pi_0( {\bf k})=\dfrac{1}{2}\frac{1}{\vert {\bf k}\vert}\;.
\end{equation}
The tree level interaction vertex (Fig. \ref{fig:Feynrules} (c)) is $\Gamma^0= -ie\gamma^0$. 
Notice that the gauge field propagator does not have the  canonical QFT dependence $(1/k^2)$. The reason is that, although the electrons are confined to the two dimensional plane, the electromagnetic field lives in three spacial dimensions. The bare propagator in (\ref{barephoton}) makes the graphene model different from QED(2+1). In particular the interaction term in the Lagrangian is scale invariant: the coupling constant of the present model is dimensionless  unlike that of QED(2+1). 

The model was shown to be gauge invariant and renormalizable and in \cite{GGV94}. The renormalization functions can be defined from the self-energy and the
vertex as :
\begin{equation}
G_0^{-1}-\Sigma(k^0,{\bf k})=Z_\psi^{-1/2}(k^0,{\bf
k})[k^0\gamma^0-Z_v(k^0,{\bf k})v\bm{\gamma} \cdot {\bf k}],
\label{zelectron}
\end{equation}
\begin{equation}
\Gamma=Z_e  e \gamma^0,
\label{zint}
\end{equation}
It is important to note that when renormalizing at a given order in perturbation theory, the vertex functions (amputated one particle irreducible Green's functions) directly related to the observable quantities have to be computed as a sum of all the corrections and counter-terms up to this order. It does not make sense to renormalize a single diagram. 

\subsection{Renormalization of the theory to first order in the interaction}
\begin{figure}
\begin{center}
\includegraphics[scale=0.5]{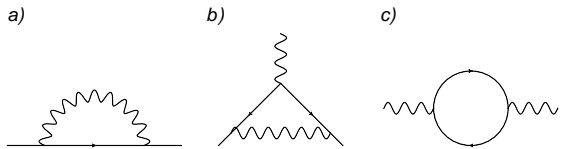}
\caption{Primitively divergent Feynman graphs in QED(4). } 
\label{fig:Primitive}
\end{center}
\end{figure}
In QED(3+1) there are three primitive divergent diagrams
shown in Fig. \ref{fig:Primitive}. 
As there are three free parameters in
the model: the electron and photon wave functions and the coupling
constant, the model is strictly renormalizable and when computing
a higher loop diagram we get higher powers of the logarithmic
divergence.

In the model of instantaneous Coulomb interaction for graphene the
only primitively divergent graph at the one loop level is the one
corresponding to the electron self-energy in Fig \ref{fig:Primitive} (a) and of this, the divergence only affects the
spacial part of the momentum. The result of the computation of the
diagram  with a hard cutoff is:
\beq
\Sigma^{(1)}_\Lambda(\mathbf{k})=-\frac{g}{4}v\bm{\gamma}\cdot\mathbf{k}
\left( -\log \frac{{\bf k}^2}{\Lambda^2}+4\log 2\right).
\label{cutoff}
\eeq
\begin{figure}
\begin{center}
\includegraphics[scale=1]{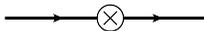}
\caption{Tree level counterterm associated to the electron self-energy. See text for details. }
\label{fig:Counterterm} 
\end{center}
\end{figure}
The electron self-energy can be made finite at this order in perturbation theory by including at tree level a counter-term of the form depicted in Fig. \ref{fig:Counterterm}  with the associated Feynman rule: 
\beq
\label{eq:eSEcountlambd}
\Sigma^{(1)}_{ct,\Lambda}(\mathbf{k})= \dfrac{g}{4}v\bm{\gamma}\cdot\mathbf{k}\left( \log\Lambda^2 + 4\log 2 + F_{\Lambda}\right),
\eeq
As expected in a renormalizable theory, the counterterm has the same operator dependence as  a term in the original Lagrangian. Since the only requirement to impose on it is to cancel the divergent part of the given diagram it contains a momentum-independent arbitrary finite part $F_{\Lambda}$ to be fixed by the renormalization condition, i. e. by an experimental measure that allows to extract the value of the  two point function at a given momentum $k_R$. This condition introduces the scale that settles the apparent dimensional mismatch in (\ref{eq:eSEcountlambd}).

Summing up the contributions of the tree level plus the two Feynman graphs of order $g$ (Fig. \ref{fig:Primitive}(a) and Fig. \ref{fig:Counterterm})  the  two-point  function is:
\begin{equation}\label{eq:InvPropInstpolmu}
G_{\Lambda}^{-1}(k^0,\mathbf{k})= -i\left( \gamma^{0}k_{0} +
v\bm{\gamma}\cdot\mathbf{k}\left[1-\dfrac{g}{4}(\log {\bf k}^2 +F_{\Lambda})\right]\right),
\end{equation}
that can be written as
\beq
G_{\Lambda}^{-1}(k^0,\mathbf{k})= -i\left( \gamma^{0}k_{0} +
v({\bf k})\bm{\gamma}\cdot\mathbf{k} \right),
\label{gren}
\eeq
with
\beq
v(\mathbf{k})=v\left[1-\dfrac{g}{4}(\log {\bf k}^2 +F_{\Lambda})\right].
\eeq

The two point function (\ref{gren}) has the same form as the free one with a k-dependent and arbitrary parameter $v(\mathbf{k})$. The last step of the renormalization program is to fix the arbitrariness with a renormalization condition. For this we need an experimental measure of  the Fermi velocity at a given value of the momentum $k_R$. 

The experimental determination of the Fermi velocity of graphene \cite{Geim09} is an important -and elusive- issue similar to the determination of the fine structure constant in QED(3+1) and we will discuss this issue further in the discussion section. As in the precisson tests of QED, each comparison between theory and experiment can be seen as an independent determination of $v_F$. To exemplify how the renormalization works we can take as an example the experimental value of the Fermi velocity given in \cite{DCetal07}: 
\beq
v(  125 {\rm meV}) = 1.093 \cdot 10^6  {\rm m/s} \equiv v_F.
\label{expv}
\eeq
With this  condition we fix the value of $F_\Lambda$ choosing the bare velocity to be $v=v_F$. The physical Fermi velocity will depend on the energy at which it is measured and on the renormalization point $k_R$  (125 meV in this case): 
\beq
v_R(\mathbf{k})=v_F\left[1-\dfrac{g}{4}\log(\frac{ {\bf k}^2}{{\bf k_R}^2} )\right].
\eeq
The definition of the running velocity defines the running  coupling constant: 
\beq
g_R(\mathbf{k})=\frac{e^2}{4\pi v_R(\mathbf{k})}.
\eeq
The coupling constant appearing in the physical magnitudes is defined as a function of the Fermi velocity as $g=\frac{e^2}{4\pi\epsilon v_F}$.
Two different renormalization prescriptions for the Fermi velocity measured at points $k_A$, $k_B$ are related by the renormalization group equation:
\beq
\frac{v(\mathbf{k}_A)}{v(\mathbf{k}_B)}= 1-\dfrac{g_{B}}{4}\log\left(\frac{\mathbf{k}_A}{\mathbf{k}_B}\right).
\eeq
Hence, it is the RG what ensures that the exact theory is independent of the experimental point chosen in (\ref{expv}). This in turn guarantees the consistency of the renormalization procedure.

The calculation in a dimensional regularization scheme follows exactly the same steps with an equivalent finite arbitrary part in the counterterm $F_\mu$ instead of $F_\Lambda$ which is eliminated  by the renormalization condition. That the procedure of renormalization does not depend on the cutoff is obvious considering that there exists a renormalization procedure (BPHZ scheme) that does not require the  use any cutoff \cite{C84}.


Notice that although being a tree level interaction, the counterterm  is of order $g$. This diagram  has to be added in the construction of the vertex functions at each given order in perturbation theory. In particular it will affect the second order diagrams computed in the next section. 


The one loop correction to the photon self-energy of Fig. \ref{fig:Primitive} (b), the polarization function, is finite and
given by
\begin{equation}
\Pi({\bf k}, \omega)=i\frac{e^2}{8}\frac{{\bf
k}^2}{\sqrt{v_F^2{\bf k}^2-\omega^2}}\;. \label{bubble}
\end{equation}
This one loop result is independent on the nature of the
interaction since only electron propagators appear in the
calculation. Different interaction vertices describing Yukawa
couplings, scalar potentials or disorder couplings may change the
tensor structure but the diagram will remain finite.  In QED(3+1) this diagram
has a logarithmic singularity and the photon polarization has higher
powers of logs at higher orders in perturbation theory what
gives rise to the electric charge renormalization.

Finally, the one loop vertex correction in Fig. \ref{fig:Primitive} (c) is
also finite  and given by
\beq
\Gamma^{0}_{(1)}(k_{0},\mathbf{k})\sim
-\dfrac{g}{4} \gamma^0 e+ \mathrm{finite}.
\eeq
The first term in this equation is the result of the explicit computation of the potentially divergent part of the diagram that turns out to be finite. The rest is encoded in the ``finite"  part that is a complicated expression of  all the  variables including Feynman parameters not important for the present discussion. 

This finishes the renormalization of the theory to first order in perturbation theory (one loop level). Any observable quantity at this level can be computed with the Feynman diagrams of Figs. \ref{fig:Feynrules} and \ref{fig:Counterterm} and will be finite and independent of the regularization procedure.   

Notice that at this order in perturbation theory there are no Coulomb interaction corrections to the optical conductivity. We will discuss the renormalization of this function to the next order in perturbation theory in the next section.

\subsection{Photon propagator to second order in perturbation theory}In most of the standard perturbative renormalizable theory like QED(3+1), second order (two loops) corrections only change slightly the values of the observables computed at first order and do not change the physics set at first order. A different case occurs when a primitively divergent diagram appears at second order, the classical example being the two loops self-energy correction in $\lambda\phi^4$ in (3+1) dimensions that gives rise to the wave function renormalization absent at the one loop level \cite{R01}. In the graphene case the new physical feature that appears at second order is related to the electron wave function renormalization whose absence at the one loop level is due to the instantaneous nature of the photon propagator. Since the main issue of the present work is the determination of observables in the renormalized theory we will focus on the photon propagator that is the most important object in the transport properties.

\begin{figure}
\begin{center}
\includegraphics[scale=0.7]{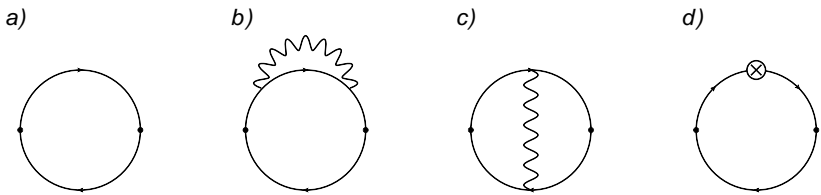}
\caption{Feynman diagrams contributing to the photon self-energy to second order in perturbation theory. }
\label{fig:bubbles} 
\end{center}
\end{figure}
The full propagator at order $g^2$ will be given by the sum of the diagrams shown in Fig.  \ref{fig:bubbles}. Diagrams  Fig. \ref{fig:bubbles}(a)  and  Fig. \ref{fig:bubbles}(c) are finite in this model. The diagram in Fig. \ref{fig:bubbles}(b) contains a sub divergent diagram  which is identified as the first order correction to the electron self-energy diagram discussed in the previous section. This divergence is canceled  by the diagram in Fig. \ref{fig:bubbles}(d) leaving a finite, unambiguous, result.

The sum of  diagrams \ref{fig:bubbles}(b) and \ref{fig:bubbles}(d) reads: 

\begin{equation}
\Pi_b+\Pi_d=-e^2 \int \frac{d^3k}{(2\pi)^3}\gamma^0 G(k)\left[\Sigma^{(1)}_{ct} -ie^2\int \frac{d^3p}{(2\pi)^3}\gamma^0 G(k+p)\gamma^0\frac{1}{|\mathbf{p}|}\right]G(k)\gamma^0 G(k+q),
\end{equation}
where $\Sigma^{(1)}_{ct}$ is given  by eq. 
(\ref{eq:eSEcountlambd}). We can identify the piece in the brackets as the renormalized electron self-energy with both contributions from the diagram in Fig.\ref{fig:Primitive}(a) and its counterterm shown in Fig.\ref{fig:Counterterm}. Hence the inner loop has already been computed, and the potentially divergent part that remains is: 

\begin{equation}
\Pi_b+\Pi_d\sim \frac{e^4}{32 \pi v_{F}} \int \frac{d^3k}{(2\pi)^3}\mathrm{Tr}\frac{\gamma^0\left(\gamma^0k_0+ v_F\bm{\gamma}\cdot\mathbf{k}\right)\bm{\gamma}\cdot\mathbf{k}\left(\gamma^0k_0+ v_F\bm{\gamma}\cdot\mathbf{k}\right)\gamma^0\left(\gamma^0(k_0+q_0)+v_F\bm{\gamma}\cdot(\mathbf{k}+\mathbf{q})\right)\log{\frac{\mathbf{k}^2}{\mathbf{k}^2_{R}}}}{(-k_0^2+v_{F}^2\bm{k}^2)^2(-(k_0+q_0)^2+v_{F}^2(\mathbf{k}+\mathbf{q})^2)},
\end{equation}

\vspace{0.3cm}
\noindent
which an explicit calculation shows to be finite. Thus, the renormalization procedure for the photon propagator is complete to second order in perturbation theory by means of the renormalization condition 
that made the electron propagator finite to first order in perturbation theory. 

The photon propagator is directly related to the optical properties of the system since it is included inside the definition of the dielectric function $\epsilon(\omega,\mathbf{q})$ as:
\begin{equation}
\frac{1}{\epsilon(\mathbf{q},\omega)}=1+V_{0}(\mathbf{q})\Pi(\mathbf{q},\omega),
\end{equation}
which in the limit $\omega \rightarrow 0$ gives the static screening properties of the system and is purely real. In undoped graphene no finite screening length is generated and the effective Coulomb  potential remains long ranged.
Both the real and imaginary parts of the photon self-energy are related to physical observables. The imaginary part of the photon self-energy is related to the frequency-dependent conductivity through:
\begin{equation}\label{cond}
\mathrm{Re}\sigma(\omega) = \lim_{\mathbf{q} \rightarrow 0} \frac{\omega}{\mathbf{q}^2}\mathrm{Im}\Pi(\omega,\mathbf{q}).
\end{equation}
From which the optical response of the system can also be obtained.
The non-interacting optical conductivity can be computed from the one-loop diagram in fig \ref{fig:bubbles}(a).  This result is well known:
\begin{equation}
\sigma_{0}(\omega) = \frac{\pi}{2}\frac{e^2}{h}.
\end{equation}
Therefore the optical conductivity at first order is a constant independent of the interaction. 
The Coulomb interaction corrections to the optical conductivity are obtained from eq. (\ref{cond}) plugging in the photon propagator at the two loops level. Since we have not computed the finite parts of the diagrams involved we can not give a precise number but from the discussion of the previous section it is clear that the result does not depend on the regularization prescription and it is determined by the chosen experimental data fixing the Fermi velocity.  

\section{Conclusions and discussion}

In this work we have addressed the determination of physical observables in
graphene in the presence of Coulomb interactions of
density-density type modeled with a static  Coulomb potential
within a QFT perturbative renormalization scheme. We have shown
that all the divergences up to two loops of the physical quantities are associated
to the one loop electron self-energy and can be regulated by a
proper renormalization of the Fermi velocity.
The consistency of the scheme presented was exemplified by the renormalization of the photon
polarization at the two loops level. We have shown that the Coulomb
interaction corrections to the optical conductivity can be fixed
unambiguously with a proper renormalization condition
(experimental input) and do not depend on the regularization
scheme. 

The QFT renormalization procedure outlined in this work shows that it can also be applied to condensed matter theories where there is an an ultraviolet physical cutoff (the inverse of the lattice spacing $\Lambda \sim 1/a$) that prevents the appearance of infinities. Although initially devised to get rid of ultraviolet divergences, the QFT renormalization in more modern approaches is a way to define the physical parameters of effective theories in a given range of energies that do not depend on the details or parameters at higher energies.

As we have seen the most important parameter in the computation of the observable quantities in graphene is the renormalized Fermi velocity that defines the Coulomb coupling constant. In the model discussed in this work the Fermi velocity grows without bound in the infrared \cite{GGV94}. This  sets a lower bound on the validity of the model that breaks down when the Fermi velocity approaches the speed of light $c$.
The scale defines an infrared cutoff $\delta$ for the theory, and can be computed as:
\begin{equation}
\delta= k_R \exp\left[-\frac{16\pi(c-v( k_R))}{e^2}\right].
\end{equation}
This estimate depends on the renormalization point $k_R$ and on $v( k_R)$. Plugging in the measured values we realize that the huge exponential suppression gives an infrared cutoff below any experimental resolution. This does not impose any real bound on the validity of the model from a experimental point of view. The limits of validity of the static model are set by internal consistency of the theory: the choice of a charge-charge interaction made to model Coulomb interactions is consistent with the static approximation since both are related to the ratio of Fermi velocity over the speed of light. When the Fermi velocity increases it would be more consistent to consider a retarded Coulomb interaction and a full interaction  vertex. The retarded model was analyzed up to one loop in \cite{GGV94} and has been revised recently in \cite{GMP10}.

For the renormalization procedure to be complete, the fine structure constant of graphene $\alpha_G = \frac{e^2}{4\pi\epsilon v_F}$ must be fixed, being and important quantity that appears in all the electronic properties of the system. A precise experimental determination is a necessity as the experiments in graphene are reaching a high degree of accuracy. The program to fix this constant can be similar to the one followed in the case of the electromagnetic fine structure constant \cite{HFG08}. In the case of graphene the theoretical determination is much simpler since there are no mass parameters relations involved in the calculations. Moreover it is very interesting the additional fact that the quantity determining $\alpha_G$ (Fermi velocity) is by itself an observable related to the one particle properties of the system. We expect that a proper combination of photoemission \cite{Zetal06,Oetal07} optical \cite{NBetal08} and transport \cite{LLA09,DSetal08} measures with the corresponding calculations in the renormalization scheme described here should be enough to determine $\alpha_G$ as precisely as needed both theoretically and experimentally as it happens in QED(3+1).

The renormalization program described in this work can be carried out to all orders in perturbation theory. As described in \cite{GGV94,GGV99}, the electron self-energy at the two loops level has a logarithmic singularity that induces a wave fuction renormalization. A new counterterm of order $g^2$ has to be added that affects the photon polarization function at the three loops level. This counterterm can be fixed by the same renormalization condition used to fix the Fermi velocity: requiring that the two point function at the two loops order has the same form as the free one, with a finite residue at $k_R$ given by a second experimental input. In the present model all physical quantities are fixed by renormalizing the electron propagator only. It is worth noticing that with the instantaneous Coulomb interaction discussed in this work, the  wave function renormalization does not give rise to an anomalous exponent and the residue of the quasiparticle in the random phase approximation done in \cite{GGV99} is finite at the Fermi surface implying that, despite the anomalous lifetime \cite{GGV96} the system is closer to a  Fermi liquid than to a marginal Fermi liquid \cite{Vetal89}.

\section{Acknowledgments}
M. A. H. V. thanks A. Gonz\'alez-Arroyo for enjoyable discussions on
renormalization in Quantum Field Theory. Support by MEC (Spain)
through grant FIS2008-00124 is acknowledged.

\bibliography{Renormalization}

\newcommand{\npb}{Nucl. Phys. B}\newcommand{\adv}{Adv.
  Phys.}\newcommand{\epl}{Europhys. Lett.}
\begin{thebibliography}{38}
\expandafter\ifx\csname natexlab\endcsname\relax\def\natexlab#1{#1}\fi
\expandafter\ifx\csname bibnamefont\endcsname\relax
  \def\bibnamefont#1{#1}\fi
\expandafter\ifx\csname bibfnamefont\endcsname\relax
  \def\bibfnamefont#1{#1}\fi
\expandafter\ifx\csname citenamefont\endcsname\relax
  \def\citenamefont#1{#1}\fi
\expandafter\ifx\csname url\endcsname\relax
  \def\url#1{\texttt{#1}}\fi
\expandafter\ifx\csname urlprefix\endcsname\relax\def\urlprefix{URL }\fi
\providecommand{\bibinfo}[2]{#2}
\providecommand{\eprint}[2][]{\url{#2}}

\bibitem[{\citenamefont{Novoselov et~al.}(2004)\citenamefont{Novoselov, Geim,
  Morozov, Jiang, Zhang, Dubonos, Gregorieva, and Firsov}}]{Netal04}
\bibinfo{author}{\bibfnamefont{K.~S.} \bibnamefont{Novoselov}},
  \bibinfo{author}{\bibfnamefont{A.~K.} \bibnamefont{Geim}},
  \bibinfo{author}{\bibfnamefont{S.~V.} \bibnamefont{Morozov}},
  \bibinfo{author}{\bibfnamefont{D.}~\bibnamefont{Jiang}},
  \bibinfo{author}{\bibfnamefont{Y.}~\bibnamefont{Zhang}},
  \bibinfo{author}{\bibfnamefont{S.~V.} \bibnamefont{Dubonos}},
  \bibinfo{author}{\bibfnamefont{I.~V.} \bibnamefont{Gregorieva}},
  \bibnamefont{and} \bibinfo{author}{\bibfnamefont{A.~A.}
  \bibnamefont{Firsov}}, \bibinfo{journal}{Science}
  \textbf{\bibinfo{volume}{306}}, \bibinfo{pages}{666} (\bibinfo{year}{2004}).

\bibitem[{\citenamefont{Novoselov et~al.}(2005)\citenamefont{Novoselov, Geim,
  Morozov, Jiang, Katsnelson, Grigorieva, Dubonos, and Firsov}}]{Netal05}
\bibinfo{author}{\bibfnamefont{K.~S.} \bibnamefont{Novoselov}},
  \bibinfo{author}{\bibfnamefont{A.~K.} \bibnamefont{Geim}},
  \bibinfo{author}{\bibfnamefont{S.~V.} \bibnamefont{Morozov}},
  \bibinfo{author}{\bibfnamefont{D.}~\bibnamefont{Jiang}},
  \bibinfo{author}{\bibfnamefont{M.~I.} \bibnamefont{Katsnelson}},
  \bibinfo{author}{\bibfnamefont{I.~V.} \bibnamefont{Grigorieva}},
  \bibinfo{author}{\bibfnamefont{S.~V.} \bibnamefont{Dubonos}},
  \bibnamefont{and} \bibinfo{author}{\bibfnamefont{A.~A.}
  \bibnamefont{Firsov}}, \bibinfo{journal}{Nature}
  \textbf{\bibinfo{volume}{438}}, \bibinfo{pages}{197} (\bibinfo{year}{2005}).

\bibitem[{\citenamefont{Zhang et~al.}(2005)\citenamefont{Zhang, Tan, Stormer,
  and Kim}}]{Zetal05}
\bibinfo{author}{\bibfnamefont{Y.}~\bibnamefont{Zhang}},
  \bibinfo{author}{\bibfnamefont{Y.-W.} \bibnamefont{Tan}},
  \bibinfo{author}{\bibfnamefont{H.~L.} \bibnamefont{Stormer}},
  \bibnamefont{and} \bibinfo{author}{\bibfnamefont{P.}~\bibnamefont{Kim}},
  \bibinfo{journal}{Nature} \textbf{\bibinfo{volume}{438}},
  \bibinfo{pages}{201} (\bibinfo{year}{2005}).

\bibitem[{\citenamefont{Jiang et~al.}(2007)\citenamefont{Jiang, Henriksen,
  Tung, Wang, Schwartz, Han, Kim, and Stormer}}]{Jetal07}
\bibinfo{author}{\bibfnamefont{Z.}~\bibnamefont{Jiang}},
  \bibinfo{author}{\bibfnamefont{E.~A.} \bibnamefont{Henriksen}},
  \bibinfo{author}{\bibfnamefont{L.~C.} \bibnamefont{Tung}},
  \bibinfo{author}{\bibfnamefont{Y.-J.} \bibnamefont{Wang}},
  \bibinfo{author}{\bibfnamefont{M.~E.} \bibnamefont{Schwartz}},
  \bibinfo{author}{\bibfnamefont{M.}~\bibnamefont{Han}},
  \bibinfo{author}{\bibfnamefont{P.}~\bibnamefont{Kim}}, \bibnamefont{and}
  \bibinfo{author}{\bibfnamefont{H.~L.} \bibnamefont{Stormer}},
  \bibinfo{journal}{Phys. Rev. Lett.} \textbf{\bibinfo{volume}{98}},
  \bibinfo{pages}{197403} (\bibinfo{year}{2007}).

\bibitem[{\citenamefont{Bostwick et~al.}(2007)\citenamefont{Bostwick, Ohta,
  Seyller, Horn, and Rotenberg}}]{Oetal07}
\bibinfo{author}{\bibfnamefont{A.}~\bibnamefont{Bostwick}},
  \bibinfo{author}{\bibfnamefont{T.}~\bibnamefont{Ohta}},
  \bibinfo{author}{\bibfnamefont{T.}~\bibnamefont{Seyller}},
  \bibinfo{author}{\bibfnamefont{K.}~\bibnamefont{Horn}}, \bibnamefont{and}
  \bibinfo{author}{\bibfnamefont{E.}~\bibnamefont{Rotenberg}},
  \bibinfo{journal}{Nature Phys.} \textbf{\bibinfo{volume}{3}},
  \bibinfo{pages}{36} (\bibinfo{year}{2007}).

\bibitem[{\citenamefont{Zhou et~al.}(2008)\citenamefont{Zhou, Siegel, Fedorov,
  and Lanzara}}]{ZSetal08}
\bibinfo{author}{\bibfnamefont{S.}~\bibnamefont{Zhou}},
  \bibinfo{author}{\bibfnamefont{D.}~\bibnamefont{Siegel}},
  \bibinfo{author}{\bibfnamefont{A.}~\bibnamefont{Fedorov}}, \bibnamefont{and}
  \bibinfo{author}{\bibfnamefont{A.}~\bibnamefont{Lanzara}},
  \bibinfo{journal}{Phys. Rev. B} \textbf{\bibinfo{volume}{78}},
  \bibinfo{pages}{193404} (\bibinfo{year}{2008}).

\bibitem[{\citenamefont{Li et~al.}(2009)\citenamefont{Li, Luican, and
  Andrei}}]{LLA09}
\bibinfo{author}{\bibfnamefont{G.}~\bibnamefont{Li}},
  \bibinfo{author}{\bibfnamefont{A.}~\bibnamefont{Luican}}, \bibnamefont{and}
  \bibinfo{author}{\bibfnamefont{E.}~\bibnamefont{Andrei}},
  \bibinfo{journal}{Phys. Rev. Lett.} \textbf{\bibinfo{volume}{102}},
  \bibinfo{pages}{176804} (\bibinfo{year}{2009}).

\bibitem[{\citenamefont{Nair et~al.}(2008)\citenamefont{Nair, Blake,
  Grigorenko, Novoselov, Booth, Stauber, Peres, and Geim}}]{NBetal08}
\bibinfo{author}{\bibfnamefont{R.}~\bibnamefont{Nair}},
  \bibinfo{author}{\bibfnamefont{P.}~\bibnamefont{Blake}},
  \bibinfo{author}{\bibfnamefont{A.}~\bibnamefont{Grigorenko}},
  \bibinfo{author}{\bibfnamefont{K.}~\bibnamefont{Novoselov}},
  \bibinfo{author}{\bibfnamefont{T.}~\bibnamefont{Booth}},
  \bibinfo{author}{\bibfnamefont{T.}~\bibnamefont{Stauber}},
  \bibinfo{author}{\bibfnamefont{N.}~\bibnamefont{Peres}}, \bibnamefont{and}
  \bibinfo{author}{\bibfnamefont{A.}~\bibnamefont{Geim}},
  \bibinfo{journal}{Science} \textbf{\bibinfo{volume}{320}},
  \bibinfo{pages}{1308} (\bibinfo{year}{2008}).

\bibitem[{\citenamefont{Li et~al.}(2008)\citenamefont{Li, Henriksenand, Jiang,
  Ha, Martin, Kim, Stormer, and Basov}}]{Letal08}
\bibinfo{author}{\bibfnamefont{Z.~Q.} \bibnamefont{Li}},
  \bibinfo{author}{\bibfnamefont{E.~A.} \bibnamefont{Henriksenand}},
  \bibinfo{author}{\bibfnamefont{Z.}~\bibnamefont{Jiang}},
  \bibinfo{author}{\bibfnamefont{Z.}~\bibnamefont{Ha}},
  \bibinfo{author}{\bibfnamefont{M.~C.} \bibnamefont{Martin}},
  \bibinfo{author}{\bibfnamefont{P.}~\bibnamefont{Kim}},
  \bibinfo{author}{\bibfnamefont{H.~L.} \bibnamefont{Stormer}},
  \bibnamefont{and} \bibinfo{author}{\bibfnamefont{D.~N.} \bibnamefont{Basov}},
  \bibinfo{journal}{Nature Phys.} \textbf{\bibinfo{volume}{4}},
  \bibinfo{pages}{532} (\bibinfo{year}{2008}).

\bibitem[{\citenamefont{Du et~al.}(2009)\citenamefont{Du, Skachko, Duerr,
  Luican, and Andrei}}]{DSetal09}
\bibinfo{author}{\bibfnamefont{X.}~\bibnamefont{Du}},
  \bibinfo{author}{\bibfnamefont{I.}~\bibnamefont{Skachko}},
  \bibinfo{author}{\bibfnamefont{F.}~\bibnamefont{Duerr}},
  \bibinfo{author}{\bibfnamefont{A.}~\bibnamefont{Luican}}, \bibnamefont{and}
  \bibinfo{author}{\bibfnamefont{E.}~\bibnamefont{Andrei}},
  \bibinfo{journal}{Nature} \textbf{\bibinfo{volume}{462}},
  \bibinfo{pages}{192} (\bibinfo{year}{2009}).

\bibitem[{\citenamefont{Bolotin et~al.}(2009)\citenamefont{Bolotin, Ghahari,
  Shulman, Stormer, and Kim}}]{BGetal09}
\bibinfo{author}{\bibfnamefont{K.~I.} \bibnamefont{Bolotin}},
  \bibinfo{author}{\bibfnamefont{F.}~\bibnamefont{Ghahari}},
  \bibinfo{author}{\bibfnamefont{M.~D.} \bibnamefont{Shulman}},
  \bibinfo{author}{\bibfnamefont{H.~L.} \bibnamefont{Stormer}},
  \bibnamefont{and} \bibinfo{author}{\bibfnamefont{P.}~\bibnamefont{Kim}},
  \bibinfo{journal}{Nature} \textbf{\bibinfo{volume}{462}},
  \bibinfo{pages}{196} (\bibinfo{year}{2009}).

\bibitem[{\citenamefont{Gonz\'alez et~al.}(1994)\citenamefont{Gonz\'alez,
  Guinea, and Vozmediano}}]{GGV94}
\bibinfo{author}{\bibfnamefont{J.}~\bibnamefont{Gonz\'alez}},
  \bibinfo{author}{\bibfnamefont{F.}~\bibnamefont{Guinea}}, \bibnamefont{and}
  \bibinfo{author}{\bibfnamefont{M.~A.~H.} \bibnamefont{Vozmediano}},
  \bibinfo{journal}{Nucl. Phys. B} \textbf{\bibinfo{volume}{424 [FS]}},
  \bibinfo{pages}{595} (\bibinfo{year}{1994}).

\bibitem[{\citenamefont{Gonz\'alez et~al.}(1996)\citenamefont{Gonz\'alez,
  Guinea, and Vozmediano}}]{GGV96}
\bibinfo{author}{\bibfnamefont{J.}~\bibnamefont{Gonz\'alez}},
  \bibinfo{author}{\bibfnamefont{F.}~\bibnamefont{Guinea}}, \bibnamefont{and}
  \bibinfo{author}{\bibfnamefont{M.~A.~H.} \bibnamefont{Vozmediano}},
  \bibinfo{journal}{Phys. Rev. Lett.} \textbf{\bibinfo{volume}{77}},
  \bibinfo{pages}{3589} (\bibinfo{year}{1996}).

\bibitem[{\citenamefont{Gonz\'alez et~al.}(1999)\citenamefont{Gonz\'alez,
  Guinea, and Vozmediano}}]{GGV99}
\bibinfo{author}{\bibfnamefont{J.}~\bibnamefont{Gonz\'alez}},
  \bibinfo{author}{\bibfnamefont{F.}~\bibnamefont{Guinea}}, \bibnamefont{and}
  \bibinfo{author}{\bibfnamefont{M.~A.~H.} \bibnamefont{Vozmediano}},
  \bibinfo{journal}{Phys. Rev. B} \textbf{\bibinfo{volume}{59}},
  \bibinfo{pages}{R2474} (\bibinfo{year}{1999}).

\bibitem[{\citenamefont{Gonz\'alez et~al.}(2001)\citenamefont{Gonz\'alez,
  Guinea, and Vozmediano}}]{GGV01}
\bibinfo{author}{\bibfnamefont{J.}~\bibnamefont{Gonz\'alez}},
  \bibinfo{author}{\bibfnamefont{F.}~\bibnamefont{Guinea}}, \bibnamefont{and}
  \bibinfo{author}{\bibfnamefont{M.~A.~H.} \bibnamefont{Vozmediano}},
  \bibinfo{journal}{Phys. Rev. B} \textbf{\bibinfo{volume}{63}},
  \bibinfo{pages}{134421} (\bibinfo{year}{2001}).

\bibitem[{\citenamefont{Khveshchenko}(2001{\natexlab{a}})}]{K01}
\bibinfo{author}{\bibfnamefont{D.~V.} \bibnamefont{Khveshchenko}},
  \bibinfo{journal}{Phys. Rev. Lett.} \textbf{\bibinfo{volume}{87}},
  \bibinfo{pages}{246802} (\bibinfo{year}{2001}{\natexlab{a}}).

\bibitem[{\citenamefont{Khveshchenko}(2001{\natexlab{b}})}]{K01b}
\bibinfo{author}{\bibfnamefont{D.~V.} \bibnamefont{Khveshchenko}},
  \bibinfo{journal}{Phys. Rev. Lett.} \textbf{\bibinfo{volume}{87}},
  \bibinfo{pages}{206401} (\bibinfo{year}{2001}{\natexlab{b}}).

\bibitem[{\citenamefont{Gusynin et~al.}(2002)\citenamefont{Gusynin, Sharapov,
  and Carbotte}}]{GMS02}
\bibinfo{author}{\bibfnamefont{V.~P.} \bibnamefont{Gusynin}},
  \bibinfo{author}{\bibfnamefont{S.~G.} \bibnamefont{Sharapov}},
  \bibnamefont{and} \bibinfo{author}{\bibfnamefont{J.~P.}
  \bibnamefont{Carbotte}}, \bibinfo{journal}{Phys. Rev. B}
  \textbf{\bibinfo{volume}{66}}, \bibinfo{pages}{045108}
  (\bibinfo{year}{2002}).

\bibitem[{\citenamefont{{Das Sarma} et~al.}(2006)\citenamefont{{Das Sarma},
  Hwang, and Tse}}]{DS06a}
\bibinfo{author}{\bibfnamefont{S.}~\bibnamefont{{Das Sarma}}},
  \bibinfo{author}{\bibfnamefont{E.}~\bibnamefont{Hwang}}, \bibnamefont{and}
  \bibinfo{author}{\bibfnamefont{W.-K.} \bibnamefont{Tse}},
  \bibinfo{journal}{Phys. Rev. B} \textbf{\bibinfo{volume}{75}},
  \bibinfo{pages}{121406(R)} (\bibinfo{year}{2006}).

\bibitem[{\citenamefont{Barlas et~al.}(2007)\citenamefont{Barlas, Pereg-Barnea,
  Polini, Asgari, and MacDonald}}]{Macetal07}
\bibinfo{author}{\bibfnamefont{Y.}~\bibnamefont{Barlas}},
  \bibinfo{author}{\bibfnamefont{T.}~\bibnamefont{Pereg-Barnea}},
  \bibinfo{author}{\bibfnamefont{M.}~\bibnamefont{Polini}},
  \bibinfo{author}{\bibfnamefont{R.}~\bibnamefont{Asgari}}, \bibnamefont{and}
  \bibinfo{author}{\bibfnamefont{A.}~\bibnamefont{MacDonald}},
  \bibinfo{journal}{Phys. Rev. Lett.} \textbf{\bibinfo{volume}{98}},
  \bibinfo{pages}{236601} (\bibinfo{year}{2007}).

\bibitem[{\citenamefont{Mishchenko}(2007)}]{Mi07}
\bibinfo{author}{\bibfnamefont{E.~G.} \bibnamefont{Mishchenko}},
  \bibinfo{journal}{Phys. Rev. Lett.} \textbf{\bibinfo{volume}{98}},
  \bibinfo{pages}{216801} (\bibinfo{year}{2007}).

\bibitem[{\citenamefont{Mishchenko}(2008)}]{Mi08}
\bibinfo{author}{\bibfnamefont{E.~G.} \bibnamefont{Mishchenko}},
  \bibinfo{journal}{Europhys. Lett.} \textbf{\bibinfo{volume}{83}},
  \bibinfo{pages}{17005} (\bibinfo{year}{2008}).

\bibitem[{\citenamefont{Kotov et~al.}(2008)\citenamefont{Kotov, Uchoa, and
  Neto}}]{KUC08}
\bibinfo{author}{\bibfnamefont{V.~N.} \bibnamefont{Kotov}},
  \bibinfo{author}{\bibfnamefont{B.}~\bibnamefont{Uchoa}}, \bibnamefont{and}
  \bibinfo{author}{\bibfnamefont{A.~H.~C.} \bibnamefont{Neto}},
  \bibinfo{journal}{Phys. Rev. B} \textbf{\bibinfo{volume}{78}},
  \bibinfo{pages}{035119} (\bibinfo{year}{2008}).

\bibitem[{\citenamefont{Herbut et~al.}(2008)\citenamefont{Herbut, Juricic, and
  Vafek}}]{HJV08}
\bibinfo{author}{\bibfnamefont{I.~F.} \bibnamefont{Herbut}},
  \bibinfo{author}{\bibfnamefont{V.}~\bibnamefont{Juricic}}, \bibnamefont{and}
  \bibinfo{author}{\bibfnamefont{O.}~\bibnamefont{Vafek}},
  \bibinfo{journal}{Phys. Rev. Lett.} \textbf{\bibinfo{volume}{100}},
  \bibinfo{pages}{046403} (\bibinfo{year}{2008}).

\bibitem[{\citenamefont{Sheehy and Schmalian}(2009)}]{SS09}
\bibinfo{author}{\bibfnamefont{D.}~\bibnamefont{Sheehy}} \bibnamefont{and}
  \bibinfo{author}{\bibfnamefont{J.}~\bibnamefont{Schmalian}},
  \bibinfo{journal}{Phys. Rev. B} \textbf{\bibinfo{volume}{80}},
  \bibinfo{pages}{193411} (\bibinfo{year}{2009}).

\bibitem[{\citenamefont{Grushin et~al.}(2009)\citenamefont{Grushin, Valenzuela,
  and Vozmediano}}]{GVV09}
\bibinfo{author}{\bibfnamefont{A.~G.} \bibnamefont{Grushin}},
  \bibinfo{author}{\bibfnamefont{B.}~\bibnamefont{Valenzuela}},
  \bibnamefont{and} \bibinfo{author}{\bibfnamefont{M.~A.~H.}
  \bibnamefont{Vozmediano}}, \bibinfo{journal}{Phys. Rev. B}
  \textbf{\bibinfo{volume}{80}}, \bibinfo{pages}{155417}
  (\bibinfo{year}{2009}).

\bibitem[{\citenamefont{Nash}(1978)}]{N78}
\bibinfo{author}{\bibfnamefont{C.}~\bibnamefont{Nash}},
  \emph{\bibinfo{title}{Relativistic Quantum Fields}}
  (\bibinfo{publisher}{Academic Press}, \bibinfo{year}{1978}).

\bibitem[{\citenamefont{Collins}(1984)}]{C84}
\bibinfo{author}{\bibfnamefont{J.}~\bibnamefont{Collins}},
  \emph{\bibinfo{title}{Renormalization}} (\bibinfo{publisher}{Cambridge
  University Press}, \bibinfo{year}{1984}).

\bibitem[{\citenamefont{Hanneke et~al.}(2008)\citenamefont{Hanneke, Fogwell,
  and Gabrielse}}]{HFG08}
\bibinfo{author}{\bibfnamefont{D.}~\bibnamefont{Hanneke}},
  \bibinfo{author}{\bibfnamefont{S.}~\bibnamefont{Fogwell}}, \bibnamefont{and}
  \bibinfo{author}{\bibfnamefont{G.}~\bibnamefont{Gabrielse}},
  \bibinfo{journal}{Phys. Rev. Lett.} \textbf{\bibinfo{volume}{100}},
  \bibinfo{pages}{120801} (\bibinfo{year}{2008}).

\bibitem[{\citenamefont{Wallace}(1947)}]{W47}
\bibinfo{author}{\bibfnamefont{P.~R.} \bibnamefont{Wallace}},
  \bibinfo{journal}{Phys. Rev.} \textbf{\bibinfo{volume}{71}},
  \bibinfo{pages}{622} (\bibinfo{year}{1947}).

\bibitem[{\citenamefont{Slonczewski and Weiss}(1958)}]{SW58}
\bibinfo{author}{\bibfnamefont{J.~C.} \bibnamefont{Slonczewski}}
  \bibnamefont{and} \bibinfo{author}{\bibfnamefont{P.~R.} \bibnamefont{Weiss}},
  \bibinfo{journal}{Phys. Rev.} \textbf{\bibinfo{volume}{109}},
  \bibinfo{pages}{272} (\bibinfo{year}{1958}).

\bibitem[{\citenamefont{Geim}(2009)}]{Geim09}
\bibinfo{author}{\bibfnamefont{A.~K.} \bibnamefont{Geim}},
  \bibinfo{journal}{Science} \textbf{\bibinfo{volume}{234}},
  \bibinfo{pages}{1530} (\bibinfo{year}{2009}).

\bibitem[{\citenamefont{Deacon et~al.}(2007)\citenamefont{Deacon, Chuang,
  Nicholas, Novoselov, and Geim}}]{DCetal07}
\bibinfo{author}{\bibfnamefont{R.}~\bibnamefont{Deacon}},
  \bibinfo{author}{\bibfnamefont{K.-C.} \bibnamefont{Chuang}},
  \bibinfo{author}{\bibfnamefont{R.~J.} \bibnamefont{Nicholas}},
  \bibinfo{author}{\bibfnamefont{.~K.} \bibnamefont{Novoselov}},
  \bibnamefont{and} \bibinfo{author}{\bibfnamefont{A.}~\bibnamefont{Geim}},
  \bibinfo{journal}{Phys. Rev. B} \textbf{\bibinfo{volume}{76}},
  \bibinfo{pages}{081406(R)} (\bibinfo{year}{2007}).

\bibitem[{\citenamefont{Ramond}(2001)}]{R01}
\bibinfo{author}{\bibfnamefont{P.}~\bibnamefont{Ramond}},
  \emph{\bibinfo{title}{Field Theory : A Modern Primer}}
  (\bibinfo{publisher}{Westview Press}, \bibinfo{year}{2001}).

\bibitem[{\citenamefont{Giuliani et~al.}(2010)\citenamefont{Giuliani,
  Mastropietro, and Porta}}]{GMP10}
\bibinfo{author}{\bibfnamefont{A.}~\bibnamefont{Giuliani}},
  \bibinfo{author}{\bibfnamefont{V.}~\bibnamefont{Mastropietro}},
  \bibnamefont{and} \bibinfo{author}{\bibfnamefont{M.}~\bibnamefont{Porta}},
  \bibinfo{journal}{arXiv:1001.5347}  (\bibinfo{year}{2010}).

\bibitem[{\citenamefont{Zhou et~al.}(2006)\citenamefont{Zhou, Gweon, Graf,
  Fedorov, Spataru, Diehl, Kopelevich, Lee, Louie, and Lanzara}}]{Zetal06}
\bibinfo{author}{\bibfnamefont{S.}~\bibnamefont{Zhou}},
  \bibinfo{author}{\bibfnamefont{G.-H.} \bibnamefont{Gweon}},
  \bibinfo{author}{\bibfnamefont{J.}~\bibnamefont{Graf}},
  \bibinfo{author}{\bibfnamefont{A.}~\bibnamefont{Fedorov}},
  \bibinfo{author}{\bibfnamefont{C.}~\bibnamefont{Spataru}},
  \bibinfo{author}{\bibfnamefont{R.}~\bibnamefont{Diehl}},
  \bibinfo{author}{\bibfnamefont{Y.}~\bibnamefont{Kopelevich}},
  \bibinfo{author}{\bibfnamefont{D.-H.} \bibnamefont{Lee}},
  \bibinfo{author}{\bibfnamefont{S.~G.} \bibnamefont{Louie}}, \bibnamefont{and}
  \bibinfo{author}{\bibfnamefont{A.}~\bibnamefont{Lanzara}},
  \bibinfo{journal}{Nature Phys.} \textbf{\bibinfo{volume}{2}},
  \bibinfo{pages}{595} (\bibinfo{year}{2006}).

\bibitem[{\citenamefont{Du et~al.}(2008)\citenamefont{Du, Skachko, Barker, and
  Andrei}}]{DSetal08}
\bibinfo{author}{\bibfnamefont{X.}~\bibnamefont{Du}},
  \bibinfo{author}{\bibfnamefont{I.}~\bibnamefont{Skachko}},
  \bibinfo{author}{\bibfnamefont{A.}~\bibnamefont{Barker}}, \bibnamefont{and}
  \bibinfo{author}{\bibfnamefont{E.~Y.} \bibnamefont{Andrei}},
  \bibinfo{journal}{Nature Nanotechnology} \textbf{\bibinfo{volume}{3}},
  \bibinfo{pages}{491} (\bibinfo{year}{2008}).

\bibitem[{\citenamefont{Varma et~al.}(1989)\citenamefont{Varma, Littlewood,
  Schmitt-Rink, Abrahams, and Ruckenstein}}]{Vetal89}
\bibinfo{author}{\bibfnamefont{C.~M.} \bibnamefont{Varma}},
  \bibinfo{author}{\bibfnamefont{P.~B.} \bibnamefont{Littlewood}},
  \bibinfo{author}{\bibfnamefont{S.}~\bibnamefont{Schmitt-Rink}},
  \bibinfo{author}{\bibfnamefont{E.}~\bibnamefont{Abrahams}}, \bibnamefont{and}
  \bibinfo{author}{\bibfnamefont{A.~E.} \bibnamefont{Ruckenstein}},
  \bibinfo{journal}{Phys. Rev. Lett.} \textbf{\bibinfo{volume}{63}},
  \bibinfo{pages}{1996} (\bibinfo{year}{1989}).

\end{thebibliography}

\end{document}